\documentclass[iop]{emulateapj}

\slugcomment{ApJ, in press}
\shorttitle{The Stellar Halo of NGC~253}
\shortauthors{Bailin et~al.}

\begin{document}

\title{The Resolved Stellar Halo of NGC 253\altaffilmark{1,2}}

\author{Jeremy Bailin\altaffilmark{3,4}, Eric F. Bell\altaffilmark{3},
	Samantha N. Chappell\altaffilmark{3}, David J. Radburn-Smith\altaffilmark{5}
	and Roelof S. de Jong\altaffilmark{6}}
\altaffiltext{1}{This paper includes data gathered with the 6.5 meter Magellan Telescopes
 located at Las Campanas Observatory, Chile.}
\altaffiltext{2}{Based on observations made with the NASA/ESA Hubble Space Telescope, obtained at the Space Telescope Science Institute, which is operated by the Association of Universities for Research in Astronomy, Inc., under NASA contract NAS 5-26555. These observations are associated with program \#10523.}
\altaffiltext{3}{Department of Astronomy, University of Michigan, 830 Dennison Building,
  500 Church Street, Ann Arbor, MI 48109}
\altaffiltext{4}{jbailin@umich.edu}
\altaffiltext{5}{University of Washington, Seattle, WA 98195}
\altaffiltext{6}{Leibniz-Institut f\"ur Astrophysik Potsdam (AIP), An der Sternwarte 16, D-14482
  Potsdam, Germany}

\begin{abstract}
We have obtained Magellan/IMACS and HST/ACS imaging data
that resolve red giant branch stars in the stellar halo of
the starburst galaxy NGC~253.
The HST data cover a small area, and allow us to accurately interpret
the ground-based data, which cover $30\%$ of the halo to a distance of $30$~kpc,
allowing us to make detailed quantitative measurements of the global
properties and structure of a stellar halo outside of the Local Group.
The geometry of the halo is significantly flattened in the same sense as the disk,
with a projected axis ratio of $b/a \approx 0.35 \pm 0.1$.
The total stellar mass of the halo is estimated to be
$M_{\mathrm{halo}} \sim 2.5\pm1.5 \times 10^9~M_{\sun}$, or $6\%$ of the total
stellar mass of the galaxy,
and has a projected radial dependence that follows a power law of index
$-2.8\pm0.6$, corresponding to a three-dimensional power law index of $\sim -4$.
The total luminosity and profile shape that we measure for NGC~253 are somewhat larger
and steeper than the equivalent values for the Milky Way and M31, but are well within
the scatter of model predictions for the properties of 
stellar halos built up in a cosmological context.
Structure within the halo is seen at a variety of scales: there is small
kpc-scale density variation and a large shelf-like feature near the middle of the field.
The techniques that have been developed will be essential for
quantitatively comparing our upcoming larger
sample of observed stellar halos to models of halo formation.
\end{abstract}


\section{Introduction}
In the favored hierarchical model of galaxy formation \citep{wr78}, stellar
halos are composed of the shredded remains of the smaller protogalactic clumps
that assembled the larger galaxy \citep{sz78,bkw01}.
At radii approaching the virial radius, the dynamical time within the halo
approaches the age of the Universe. Stellar halos thus retain both
spatial and kinematic information about this assembly history
that is otherwise difficult to obtain.

The stellar halo of our own galaxy has been extensively surveyed.
Studies of high velocity
stars in the solar neighborhood and bright stars at large distances have revealed
a halo that follows a power law density distribution, a total luminosity of
approximately $10^9~L_{\sun}$, and abundant spatial, kinematic, abundance,
and stellar population
substructure \citep[e.g.][]{helmi-etal99,majewski-etal03,yanny-etal03,belokurov-etal07,
carollo-etal07,bell-etal08,bell-etal10,xue-etal10,cooper-etal11}.
However, our position within the disk of the
Milky Way makes it difficult to obtain a clear picture of the global structure
of the halo. This is additionally complicated by the fact
that different tracer populations
must be used to determine the structure at different distances.

Some information has been obtained about the halos of external galaxies by
searching
for surface brightness features in deep imaging \citep[e.g.][]{beck-etal82,
malin97,zheng-etal99,martinezdelgado-etal08,martinezdelgado-etal10}.
These observations have identified prominent
tidal streams that buttress the hierarchical model of galactic halo formation.
However, the fundamental limits in such studies are flat fielding errors and
the brightness of the background sky, which make global halo properties and
very faint surface brightness features inaccessible \citep{dejong08}.

In order to overcome these difficulties, a new technique has recently
been employed: resolving individual bright stars,
particularly red giant branch (RGB) stars which are present in all stellar
populations,
in the halos of external galaxies.
Within the past few years, knowledge of the halo of M31 has blossomed
with this technique, including the discovery of a metallicity gradient,
rich spatial and kinematic substructure, and a global profile that is
remarkably similar to that of the Milky Way
\citep{chapman-etal06,kalirai-etal06,ibata-etal07,mcconnachie-etal09,tanaka-etal10}.
Detections of halo stars in more distant galaxies are now beginning
to be achieved, such as in individual small HST/ACS fields of
NGC~5128 \citep{rejkuba-etal05} and M~81 \citep{durrell-etal10},
and in NGC~891 using
ground-based Subaru/Suprime-Cam data over a large field of view
\citep{mouhcine-etal10}.

One particularly interesting target is NGC~253. This
nearly edge-on starburst galaxy,
one of the largest galaxies in the ``Sculptor group'' \citep[which is not
truly a bound group, but rather a filament extended along the line of
sight;][]{jerjen-etal98,karachentsev-etal03-scl}, is of similar luminosity
to the Milky Way and M31, allowing direct comparisons. Deep optical images
have long shown an optical halo extending beyond its disk, with evidence for a shelf-like
feature to the south of the disk \citep{beck-etal82,malin97}.
\citet{fitzgibbons90}, in a photographic survey of the galaxy, also
detected extended structure to the south of the galactic disk, particularly
in the $I$ band. He found a total light profile in the halo that declines
as a power law with an index of $\sim -2.5$ and $\sim -3.5$ in the $V$ and
$I$ bands respectively.

Resolved near-infrared stellar photometry of the south-east quadrant
of NGC~253 was performed by \citeauthor{davidge10-n253} (\citeyear{davidge10-n253};
see also \citealp{comeron-etal01}, who detected blue main sequence stars
in the halo near the minor axis of the galaxy, most likely associated with
the outflow from the central starburst). He detected an extended flattened
distribution of asymptotic giant branch (AGB) stars
extending out to $13$~kpc from the disk plane.
Based on the similarity of the disk and halo $K$-band luminosity functions,
which implies an extended star formation history, he concluded that
these stars were deposited into the halo via the disruption of the galactic
disk in an interaction.
The data were not of sufficient
depth to detect RGB stars, which would be far more numerous and trace
stellar populations of all ages.

Theoretical models of galaxy formation in a $\Lambda$CDM universe
have been used to predict the properties of stellar halos, usually with
the assumption that the stars that are today found in the halo were
accreted in the form of satellite galaxies that were subsequently tidally
shredded \citep{bkw01,bj05,purcell-etal07,cooper-etal10}. These models predict that
halos contain several percent of the total stellar mass of the galaxy,
and have very clumpy structure, particularly in the outer regions. They
also predict significant galaxy-to-galaxy scatter in the total size and
morphology of the halo structure \citep{bj05,bell-etal08,johnston-etal08}.
While these properties qualitatively
match observed galactic halos, quantitative measurements of a larger
sample of halos is required in order to adequately test these models
and distinguish between them.

Our goal is to test these models by obtaining resolved data of stellar halos
in a larger, and therefore more representative, sample of galaxies.
This will allow us to accurately determine stellar halo structural properties,
and to investigate their correlation with the global parameters of the associated
galaxies. The larger sample will help determine any scatter between systems due
to a stochastic formation history.
Because of the large degree of predicted and observed
substructure, obtaining an unbiased view of the halo requires a large field of view
that is only possible using ground-based data. However, identification and
accurate photometry of stars in galaxies beyond the Local Group is far
superior with HST. We are therefore using a combination of HST and
ground-based wide-field imaging.

In this paper, we present the first results of this programme, a study
of NGC~253, along with a detailed description of the analysis required
to turn the data into quantitative global measures of halo structure.
We adopt a distance modulus to NGC~253 of $m-M=27.71$ ($D=3.48$~Mpc) throughout,
as determined using GHOSTS \citep{ghosts11},
and consistent with but towards the upper end of distance estimates based on
a variety of methods catalogued in the NASA Extragalactic Database%
\footnote{http://nedwww.ipac.caltech.edu/} (NED).
Note that we use the term ``halo'' to refer to the stellar halo of the galaxy,
not the dark matter halo, unless specifically stated otherwise.

\section{Data}

\subsection{Observations and Reduction}

\subsubsection{Magellan/IMACS Data}
The ground-based data were obtained at the Magellan 1 (Baade) 6.5m telescope
using the Inamori-Magellan Areal Camera \&\ Spectrograph (IMACS),
a mosaic of $8$ 2k x 4k CCDs, in f/2 configuration. This results in
a $0\farcs2$/pixel plate scale.
The field was centered $14\arcmin$ south of the center of
NGC~253, and therefore included a significant fraction of the galactic
disk within the $24\arcmin$ unvignetted field of view.
At the assumed distance of NGC~253, the field extends $30$~kpc from
the galaxy center.
On the nights of 17 -- 18 October 2009, we obtained
$1800$ seconds in CTIO $I$-band and $2650$ seconds in Bessel $V$-band. A further
$1000$ seconds were obtained in $V$-band on the night of 20 May 2010.
The seeing full width half max (FWHM)
was approximately $1\farcs0$ -- $1\farcs2$ during the exposures.
Landolt photometric standards were taken on these and other nights during
the same runs.

\begin{figure}
\plottwo{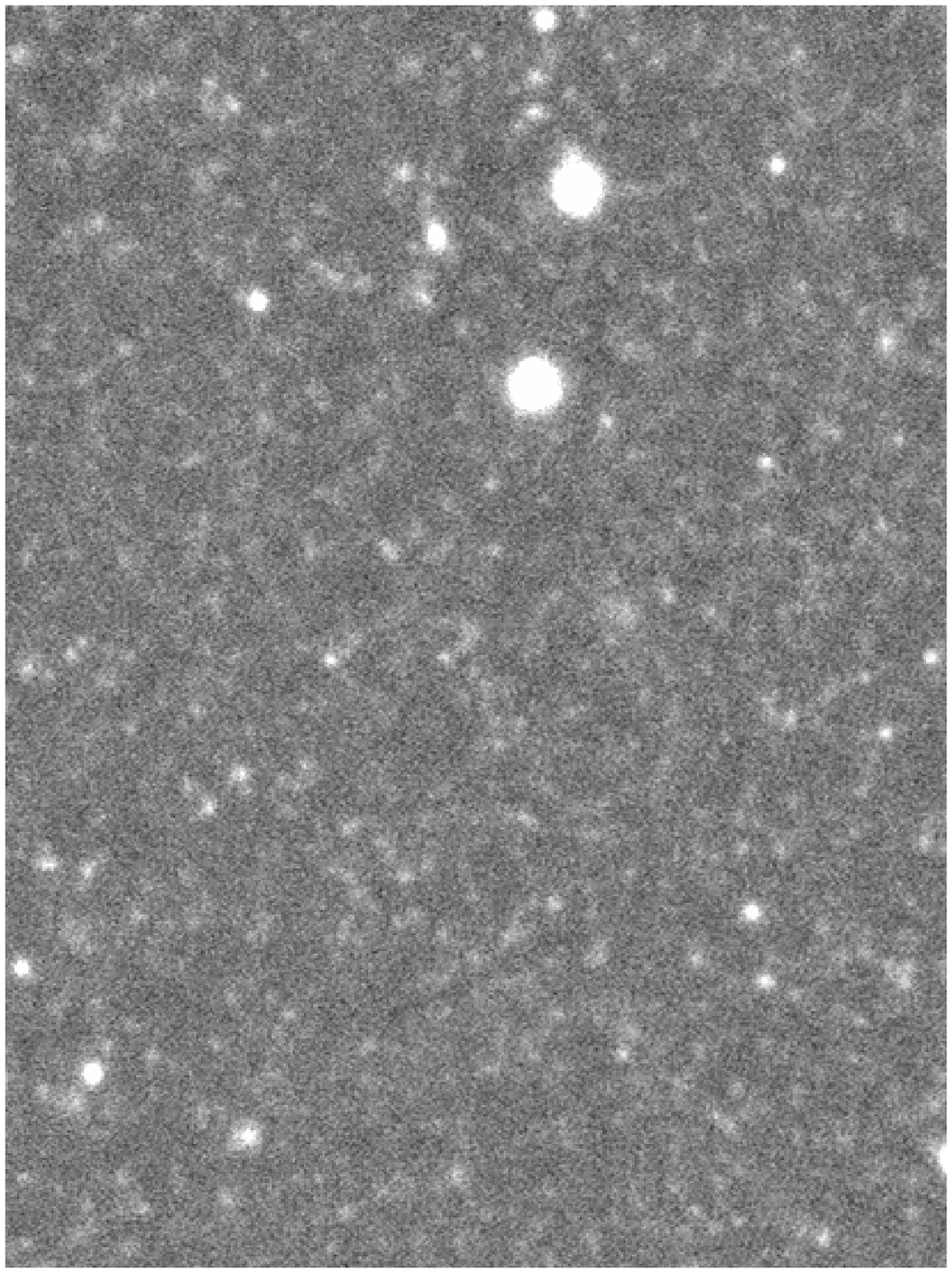}{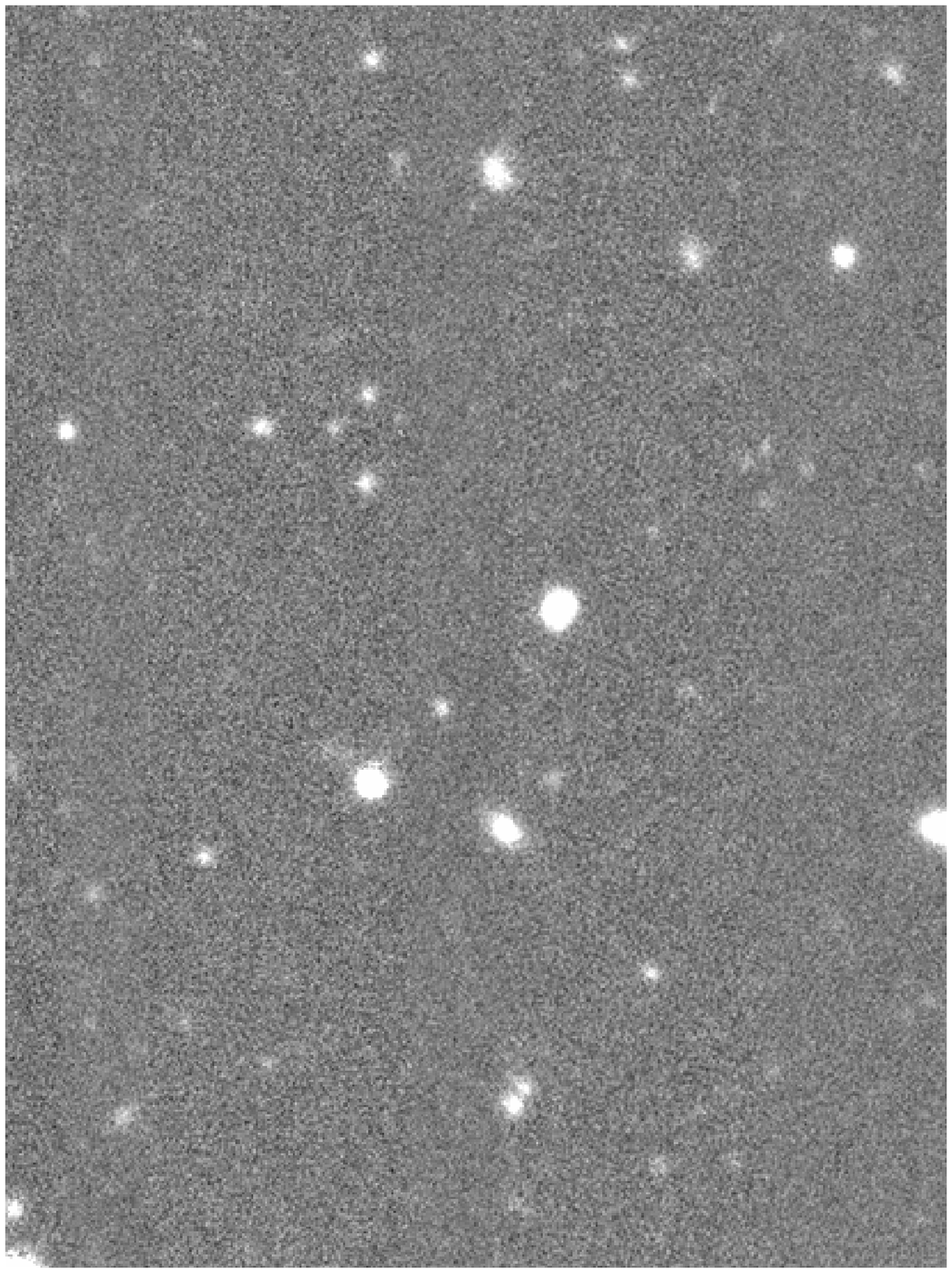}
\caption{\label{fig:field}%
\textit{Left:} Sky-subtracted stacked IMACS
$I$-band image of a $82\arcsec \times 110\arcsec$
region located $\sim 9\farcm5$ ($10$~kpc) from the center of NGC~253, along the minor axis.
RGB stars within its stellar halo are clearly visible as faint point sources.
\textit{Right:} As in the left panel, for a region located $\sim 22\arcmin$ ($22$~kpc)
from the center of NGC~253. Very few if any RGB candidates are visible.}
\end{figure}

Data were reduced using standard IRAF routines%
\footnote{IRAF is distributed by the National Optical Astronomy Observatory, which is operated by the Association of Universities for Research in Astronomy (AURA) under cooperative agreement with the National Science Foundation.},
assisted by scripts developed for reducing Maryland-Magellan Tunable Filter
data\footnote{http://www.astro.umd.edu/$\sim$veilleux/mmtf/datared.html}.
Note that although the IMACS documentation suggests using the column and row
overscan regions instead of bias frames, we found that in the case of
broadband imaging, saturated stars near the edges of the chips can bleed
into the overscan regions, making them unusable. We therefore used bias frames
that were obtained during the runs.
Sky subtraction was complicated by the presence of the galactic disk in the field,
which made it impossible to fit an accurate low-order surface to the sky.
We therefore determined the sky level using a two-dimensional
median filter of width $101$ pixels; this width was chosen to be large enough
that photometry tests on individual stars found no bias, but small enough that
the sky was flat to within $1\%$ on large scales. This results in good
image quality over most of the frame, but clearly invalidates the portion
of the field that includes the disk of NGC~253, which contains smooth features
on scales much larger than the filter width. However, as the goal of this
work is to study the halo of the galaxy, this does not affect our results.
An astrometric solution over the field was computed using the USNO~A2.0 catalog
\citep{usno-a2}.
A third order polynomial solution was required in order that the astrometric
residuals did not vary systematically across the field;
the resulting solution achieved
an rms scatter of $0\farcs24$, comparable to the pixel size.
Two sample regions of the final stacked and sky-subtracted $I$-band image
are shown in Figure~\ref{fig:field}: one located $\sim 9\farcm5$ ($10$~kpc) from the
center of the galaxy and one located at $\sim 22\arcmin$ ($22$~kpc), both approximately
along the minor axis. The faint point sources that are abundant in the inner
field but almost completely absent in the outer field are RGB stars in the halo
of NGC~253.
We ignore the very outermost part of the mosaic, which is compromised by
the shadow of the guide star camera, whose location varied between
exposures.

\subsubsection{GHOSTS}
We use data from the HST/GHOSTS survey \citep{dejong-etal07,ghosts11}
overlapping with our IMACS pointing for the purposes of
carrying out our artificial star tests, and to test our inferred stellar
densities (in the high stellar density regime).  We use HST/ACS F606W and
F814W data for NGC~253's Fields 9 and 10 \citep{ghosts11}, with
exposure times of $2 \times 680$s in both F606W and F814W bands.  Data were
reduced and calibrated using the standard HST data reduction,
and photometry was performed using the ACS module
of DOLPHOT\footnote{http://purcell.as.arizona.edu/dolphot/}.
Photometric parameters were converted to $V$ and $I$-band
magnitudes using the conversions presented in \citet{sirianni-etal05}.
Typical uncertainties (random and systematic) are $<0.05$ mag at the $I<26$
limits of interest in this paper.

\subsection{Photometry}
Sources were found on the stacked IMACS $I$-band image using the IRAF DAOFIND task,
with a threshold of $3$ times the typical standard deviation of the background.
Aperture photometry was performed using the
IRAF APPHOT package on the stacked $V$- and $I$-band images at the locations
found by DAOFIND, using an aperture of radius $8$ and $6$ pixels respectively,
matched to the width of the point spread function (PSF)
to maximize the signal-to-noise.
Aperture corrections were calculated in each band using 5 bright unsaturated
relatively isolated stars.
Magnitudes were de-reddened using the Galactic
foreground extinction values catalogued in NED for NGC~253.
The final catalogue consists of all stars that have
a positive measured flux in both bands.

Photometric calibration was performed using aperture photometry of standard
stars in the Landolt fields SA98, SA101, and SA114 taken on the nights of
17 -- 19 October 2009,
using an aperture of 40
pixels, sufficient to enclose the entire flux of each star while not
overlapping with neighboring stars. We fit a zero point, as well as airmass, color
and radial terms\footnote{The radial term accounts for the increase in solid angle subtended
by pixels at larger radii.} to the residuals. The airmass term in the $I$-band
and the color terms in both bands were found to be negligible, and so were
assumed to be zero. The resulting photometry had an rms dispersion of 
$0.051$~mag in $I$-band and $0.054$~mag in $V$-band,
constant over the $10 < V < 15.3$ magnitude range of the standard stars.
As will be seen later, such a level of photometric calibration accuracy
is sufficient for our purposes, as it is smaller than other sources of
systematic uncertainty.

\begin{figure}
\plotone{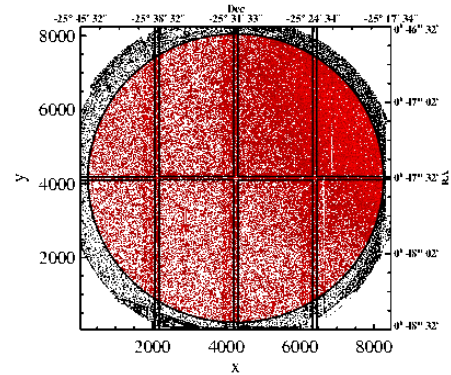}
\caption{\label{fig:allsources}%
The locations of all stellar sources are plotted in pixel coordinates on the
bottom-left axes and angular coordinates on the sky on the upper-right axes.
Points within the circle are considered far enough from the edge of the
field that they are usable (usable points are colored red in the online
version of this figure). Outlines of the footprint of the IMACS chips for one particular
dither position are shown by the rectangles.}
\end{figure}

A plot of the locations of all sources is shown in Figure~\ref{fig:allsources}.
The points within the circle were used;
this circular boundary is also used in the Monte Carlo
simulations discussed in Section~\ref{sec:densitycalib} to determine the effective
surveyed area.
Although the dithering pattern fully covered the chip gaps, the effective exposure time in
the areas covered by chip gaps is lower and therefore the noise level is higher.
This is reflected in the larger number of counts above the threshold
in those regions, since the fixed threshold is no longer equal to $3\sigma$.
However, the excess counts are predominantly at the faint end,
fainter than the brightness of the RGB candidates.
The impact of the chip gaps is therefore small, although it is large enough that
artificial structure along the horizontal chip gap is faintly visible in maps of RGB stars.

\section{Color-Magnitude Diagrams and Artificial Star Tests}

In this section, we present the color-magnitude diagrams (CMDs) of the
ground-based and HST/ACS data, identify RGB stars in the data,
and use artificial star tests to understand
the relationship between the RGB star counts derived using the ground-based data
and the intrinsic stellar population.

\subsection{Color-Magnitude Diagrams}

\begin{figure*}
\plottwo{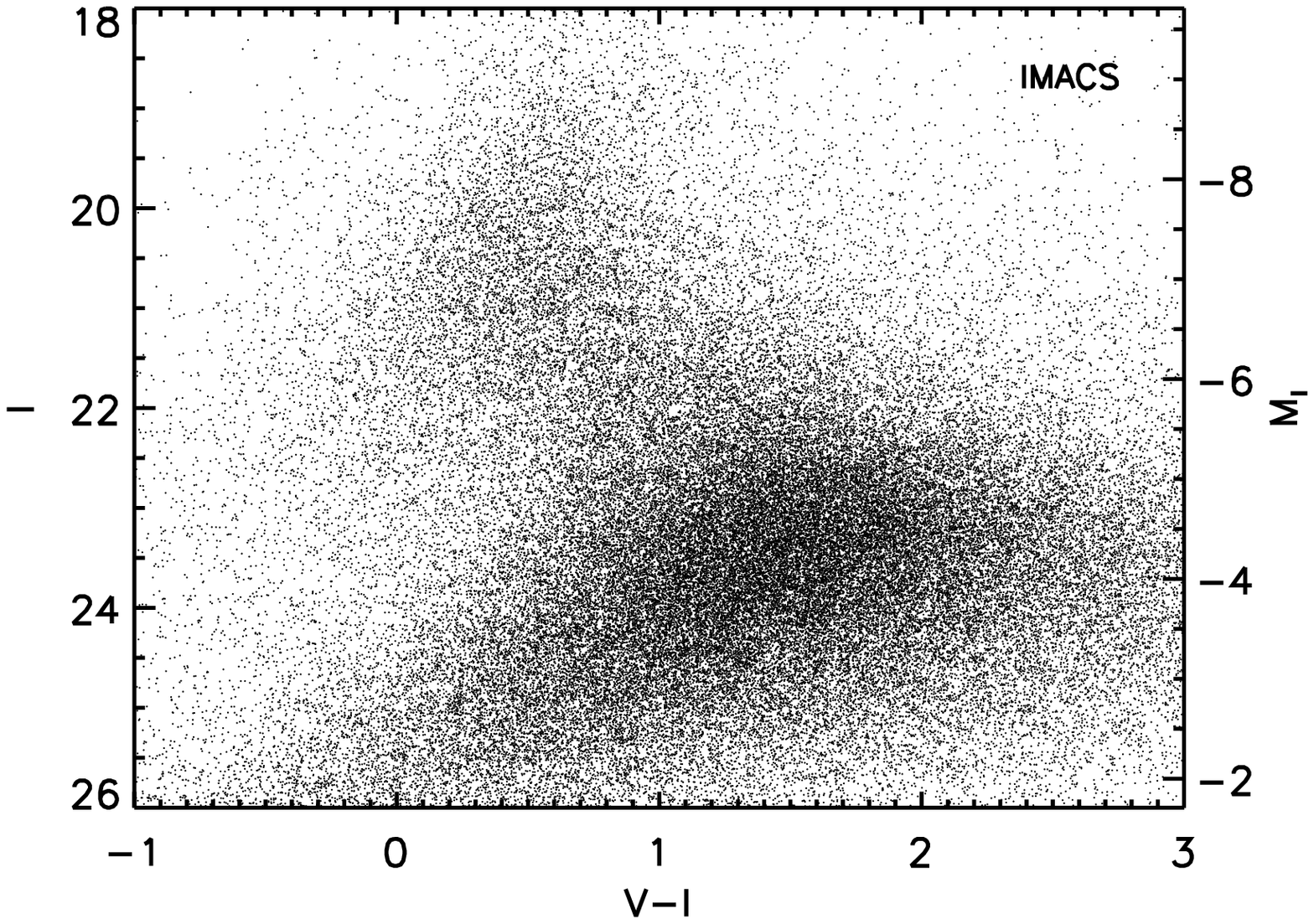}{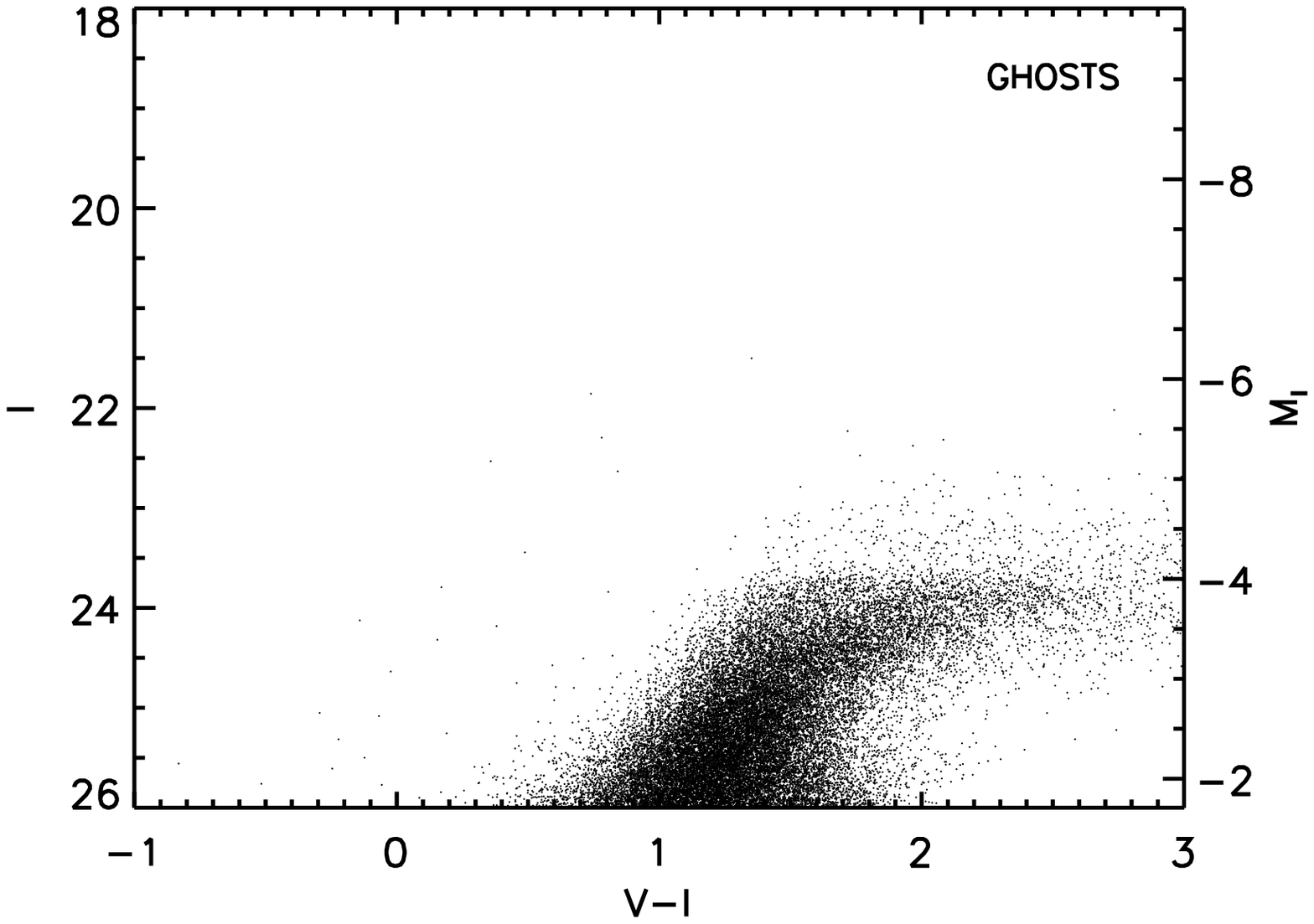}
\caption{\label{fig:cmd}%
Color-magnitude diagrams of the Magellan/IMACS field \textit{(a)} and
GHOSTS HST/ACS field 10 \textit{(b)} around NGC~253. The absolute magnitude
scale on the right-hand axis assumes a distance modulus of $m-M=27.71$.}
\end{figure*}

In Figure~\ref{fig:cmd}a, we present the CMD of all 
recovered point sources within the main IMACS
field. The bright blue vertical feature is primarily composed of
pieces of the main body of NGC~253 that have been selected as
stars by DAOFIND, along with the shredded diffraction pattern of
foreground Milky Way main sequence stars, and background disk galaxies.
At the bottom, a wide feature that appears to be the RGB is seen.
However comparison with the GHOSTS Field 10 data (Figure~\ref{fig:cmd}b) reveals that
this putative RGB extends to much brighter magnitudes than in the HST/ACS data.
Comparison of the IMACS and HST catalogs over their area of overlap reveals that
the majority of ``stars'' detected in the IMACS catalog are blends of two or more 
stars. Therefore, while the detections in IMACS are indeed detections of RGB stars
in the halo of NGC~253, interpretation of the number counts requires artificial
star tests in order to determine the relationship between the numbers and properties
of stars in the IMACS catalog and the true numbers and properties of the stellar
population.

\subsection{Artificial Star Tests}
In order to test the photometric accuracy and recovery/contamination rates
of TRGB stars in NGC 253, we carried out a suite of artificial star tests. 
Artificial stars with $m_I < 26$ were placed into 
a 1k x 1k region in the outer parts of the
$V$ and $I$-band NGC~253 images with empirical PSFs generated from
well-exposed stars.  $V$-band and $I$-band apparent magnitudes were
Monte-Carlo sampled from the CMD of the HST/GHOSTS NGC~253 Field 10
pointing.  The stars were distributed
randomly across the region at 8 different densities
(labelled ``30k'' through ``4000k'' depending on the number
of artificial stars that were added), spanning the observed
range of RGB star densities seen in the IMACS image and in the
GHOSTS data. The images were then output
and processed by the photometric pipeline in the same way as the data.

\begin{figure*}
\plotone{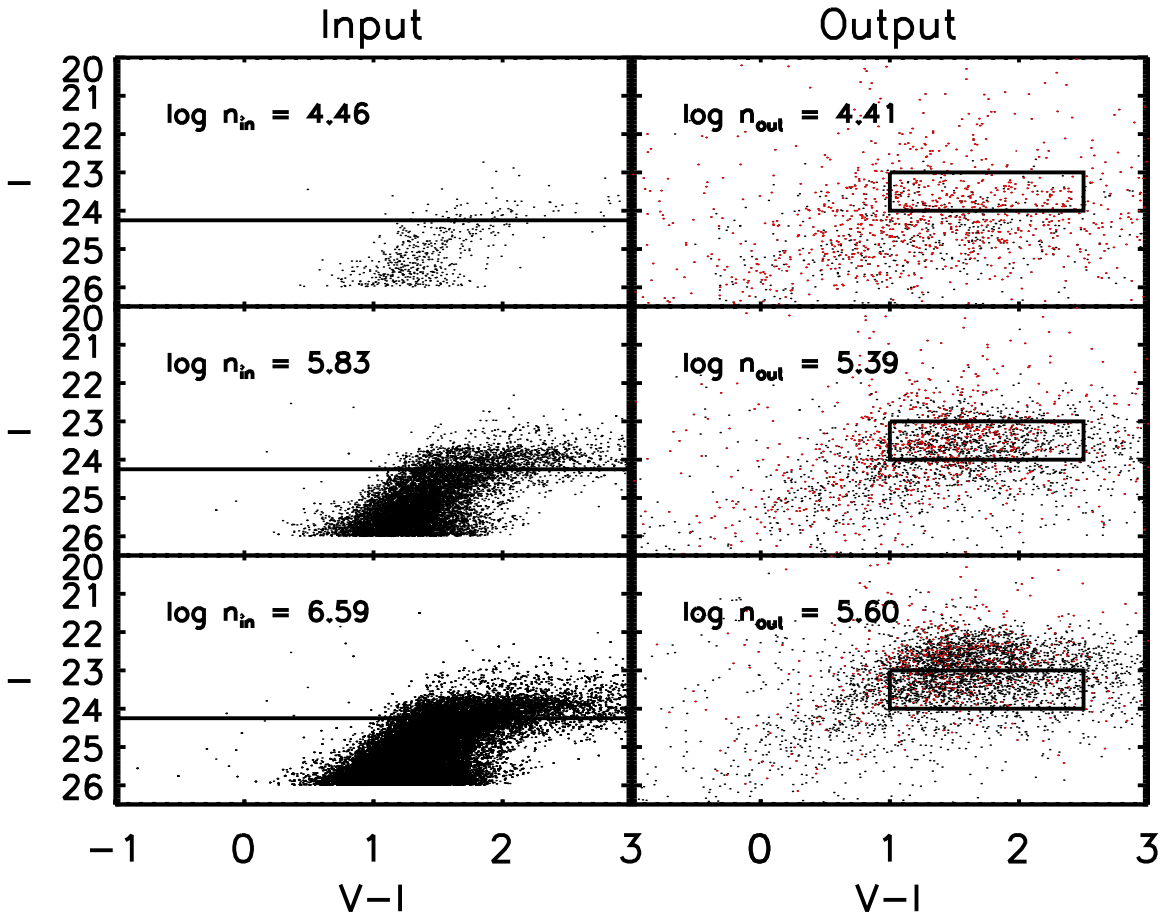}
\caption{\label{fig:artcmd}%
Artificial star CMDs. The left column shows input magnitudes of the artificial stars
that were placed in the artificial star frames, while the right column
shows the CMDs determined by performing photometry on
the output artificial star frames. In the output
CMDs, red points are those located within $0\farcs 5$ of a star in
the original image.  The rows correspond to injecting $30$k \textit{(top)},
$700$k \textit{(middle)}, and $4000$k \textit{(bottom)}
artificial stars. The input and output density of stars is
given in each panel, in number per square degree.
The lines in the left column and the boxes in the right column
denote the TRGB magnitude threshold and the boundaries of
the RGB selection box respectively.}
\end{figure*}

\label{sec:densitycalib}%
Photometry was performed on the artificial star frames exactly as for the
original frames.
CMDs generated from the input magnitudes and from the output photometry on the
artificial star frames for three sample artificial star tests
(the $30$k, $700$k and $4000$k tests) are shown in Figure~\ref{fig:artcmd}.
As the stellar density increases, the RGB stars
become blended and are detected at magnitudes significantly brighter than
those of individual stars.
This results in a CMD exactly like the one observed if the majority of
our sources are found in regions of high density.

In order to determine the number density of stars at the tip of the red giant
branch (TRGB), we have compared the number of input RGB stars above a
threshold of $I < 24.25$ (corresponding to an absolute magnitude of
$M_I < -3.46$ at the distance of NGC~253) to the number density of output
detections within a box extending from $I=23$ to $24$, and from $V-I=1.0$
to $2.5$. This box lies entirely above the input threshold in order to better capture
the bulk of the halo stars at the typical densities of our data
(see Figure~\ref{fig:artcmd}).
We have tested several different criteria for both the TRGB threshold and
the magnitude boundaries of the selection box, ranging by $\pm 0.5$~mag.
We find that this introduces
$\sim 20\%$ changes in the derived number densities once they are normalized
by the expected number of stars above each TRGB threshold for an
old metal-poor population, as discussed in
Section~\ref{sec:haloluminosity}.

\begin{figure}
\plotone{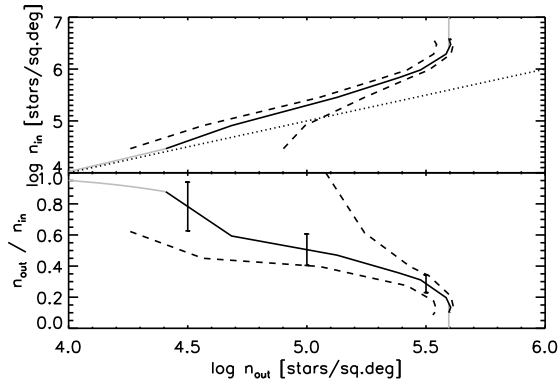}
\caption{\label{fig:artrecov}%
Results of the artificial star tests. The $x$-axis gives the density of stars
recovered within the RGB selection box in the artificial star frame, while the
$y$-axis gives the density of artificial stars injected into the artificial
star frames that lie above the TRGB threshold \textit{(top)},
and the ratio of output to input stars
\textit{(bottom)}.
The dotted line in the upper panel denotes $n_{\mathrm{in}} = n_{\mathrm{out}}$.
The right dashed line shows the case where no correction
is made for the presence of stars in the baseline image. Subtracting the density of stars
found in the original image gives the left dashed line, while subtracting only
the stars in the original image that are located within $0\farcs 5$ of a star
in the artificial star frame gives the adopted solid line. The gray lines
show the adopted extrapolations.
The error bars in the bottom panel indicate the $20\%$ uncertainty due to
the choice of a particular TRGB magnitude threshold and CMD selection box.}
\end{figure}

The relationship between the input number density above the TRGB threshold
and the output number density within the selection box is shown as the right-hand
(high-density) dashed line in the upper panel of Figure~\ref{fig:artrecov},
while their ratio (i.e. the recovered fraction) is shown in the bottom panel.
A problem is immediately apparent:
because there are $183$ detections within the RGB selection box in the ``blank field'' to which
the artificial stars were added (a combination of background galaxies and real
RGB stars in the stellar halo), more stars are detected than were added at
low densities. This is a property of the baseline image, not of our ability
to detect stars, and so is artificial. One could simply subtract $183$ from the
number of output stars (shown as the left-handed dashed line
in Figure~\ref{fig:artrecov}), but this is also incorrect: at high densities, we
approach the confusion limited regime, and the original detections have mostly been
covered up by artificial stars and are therefore no longer detected.
A better approach is to compare the locations of the detections in the artificial
star frames to the locations of the original detections in the blank field.
We performed a globally-optimized one-to-one match of the closest original
detection to each detection in the artificial star frame, and omitted
any whose best match was within $0\farcs 5$. This results in the black solid
line in Figure~\ref{fig:artrecov}. The fraction of blank field sourcees
that are omitted,
$f_{\mathrm{bg,omit}}$, is a useful measure of the degree to which individual
sources are likely to be covered up.

At the confusion limit, the number of output detections saturates, and
even begins to decrease as the majority of the blends become brighter
than the selection box. We therefore set a threshold of
$\log n_{\mathrm{out}}=5.60$, at which we can only set a lower limit on the
true density ($\log n_{\mathrm{in}} \ge 6.39$), shown as the right-hand vertical
gray line in Figure~\ref{fig:artrecov}. Fortunately, very little of the
data reach this limit. At the low density end, we note that the recovered
fraction approaches unity, and so we make the ansatz that the recovered
fraction has the form of a decaying exponential at densities below that of
our $30$k artificial star test; this results in the gray line
extending to the left, which we use to extrapolate to lower densities.
Again, most of our data lie above the point where we begin using
this extrapolation, at $\log n_{\mathrm{out}} \le 4.39$
($\log n_{\mathrm{in}} \le 4.46$).
Our final correction from the artificial star test is therefore the combination
of the solid black and gray lines in the top panel of Figure~\ref{fig:artrecov}.
Interpolation is performed linearly between $\log n_{\mathrm{out}}$ and
$\log n_{\mathrm{in}}$. The error bars in the bottom panel indicate the
$20\%$ uncertainty due to the choice of TRGB threshold magnitude and 
CMD selection box.

Note that in order to convert number counts to densities, we required the effective
area of the observed region. This is also required in order to calculate
the densities in annuli, sectors, pixels in a density map, and the footprints
of the GHOSTS pointings.
Whenever the effective area was required, Monte
Carlo simulations were performed. $N_{\mathrm{MC\,tot}}=10^7$ points were randomly sampled
uniformly over a patch of the unit sphere of solid angle
$\Omega_{\mathrm{MC\,tot}} = 0.2088\sq\degr$
encompassing the entire IMACS field. Within any given region, we counted the number
of Monte Carlo points lying within that also lie within the field where stars
were chosen (i.e. within the circle on Figure~\ref{fig:allsources}),
$N_{\mathrm{MC}}$, and also the number of bad pixels,
$N_{\mathrm{bad}}$. The bad pixels included not only pixels formally counted as bad
during the data reduction, but also regions that were completely devoid of detections due to
bleed trails or the presence of a bright foreground star. We assume that the effective
solid angle of each bad pixel is $\Omega_{\mathrm{bad}}=0.04\sq\arcsec$,
as given by the formal IMACS f/2
plate scale; the true effective area varies over the field, but this effect is small
relative to the two order of magnitude variation in the number density of RGB stars
across the field, particularly given that the fraction of bad pixels is small.
The total solid angle of the region is then given by
\begin{equation}
  \Omega_{\mathrm{region}} = N_{\mathrm{MC}} \frac{\Omega_{\mathrm{MC\,tot}}}
    {N_{\mathrm{MC\,tot}}} - N_{\mathrm{bad}} \Omega_{\mathrm{bad}}.
\end{equation}

\subsection{Background Galaxy Subtraction}
Some fraction of the detections must be due to background galaxies. To assess
the impact of this background, we have used a
control field that was created using $V$ and $I$-band imaging data taken
using the Wide Field Imager on the ESO/MPG La Silla 2.2-m telescope of the
Extended Chandra Deep Field South \citep[see][]{gawiser-etal06}.
These calibrated images cover a slightly larger field of view than the
IMACS data, and are somewhat deeper and have a smaller point spread function ($0\farcs9$).
The data were resampled onto  $0\farcs2$ pixels (to match IMACS), convolved with
the correct Gaussian to bring the FWHM to $1\farcs04$ and $1\farcs2$ for $I$-band and
$V$-band respectively, rescaled to the same zero point as the IMACS data,
and sky noise was added to bring the RMS of the image into agreement with
the IMACS data for NGC~253.  We then ran the images through our pipeline
for star detection and photometry, and used the output photometric catalog
as our ``control field''.
The density of sources within the RGB selection box in the control field is
$\log n_{\mathrm{control}}=4.27~\sq\degr^{-1}$. We scale this density
by $f_{\mathrm{bg,omit}}$ before subtracting it; at densities below
that of the $30$k artificial star frame we subtract the full
control density.
The background counts are much smaller than the densities
we derive over much of the halo,
but are comparable to the densities in the outermost regions of the field.

\section{Halo Structure}

\subsection{Profiles}\label{sec:profiles}

\subsubsection{Radial Profile}
\label{sec:radprofile}

\begin{figure}
\plotone{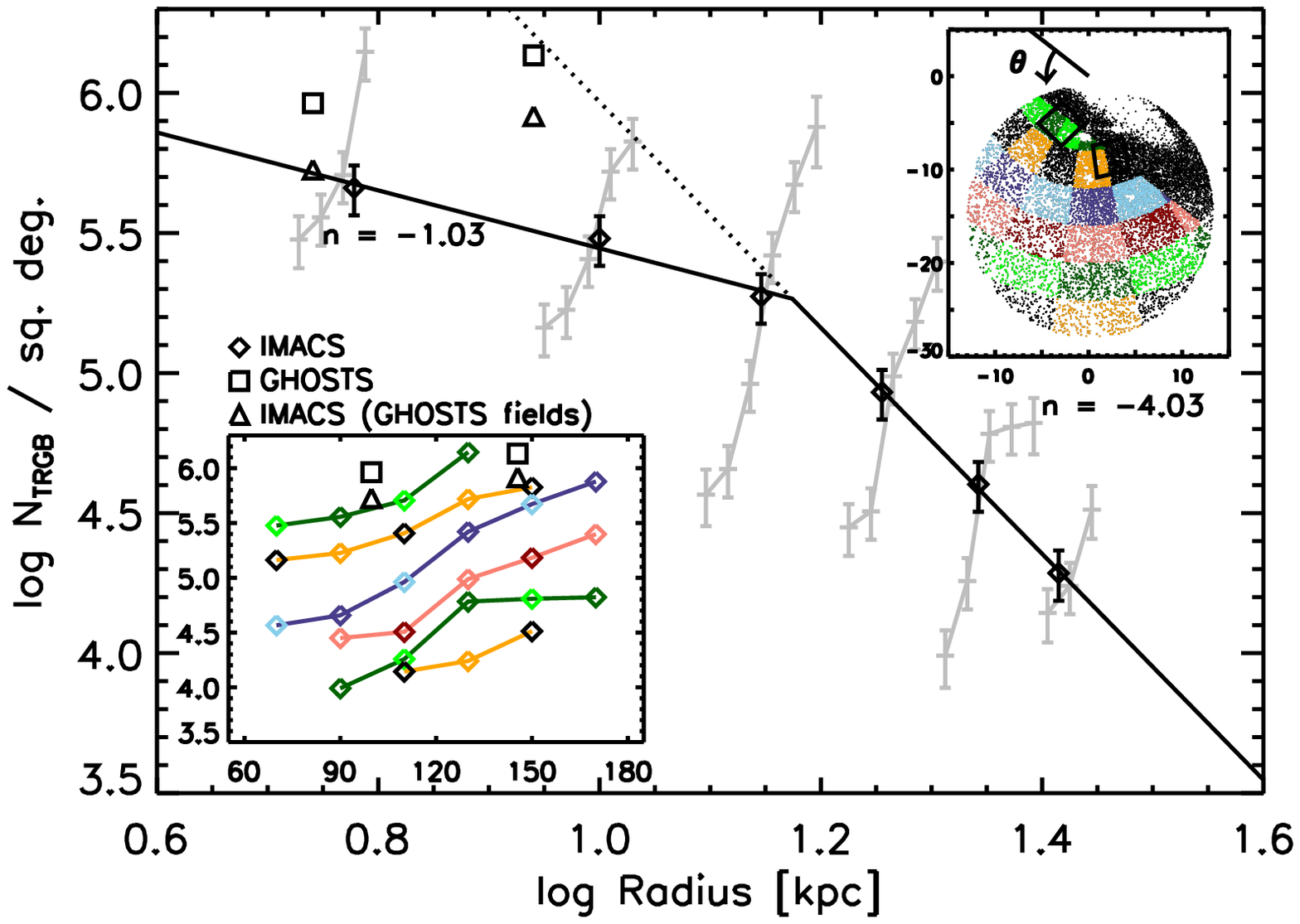}
\caption{\label{fig:radsectors}%
\textit{Main panel:}
The radial profile of the projected TRGB stellar density in the IMACS field, as defined in
Section~\ref{sec:densitycalib}. The diamonds indicate the
average value within each radial annulus. The dark solid line is a double
power-law fit to the radial profile, which has a break at $14.9$~kpc. The continuation
of the outer power-law to smaller radii is shown as the dotted line.
The TRGB number densities found in the HST/ACS GHOSTS data are denoted by the squares,
while the densities found in the same fields using IMACS data are shown as the triangles.
The azimuthal profile within a given radial annulus is shown as the gray
line at each radius.
The plotted error bars for both profiles account for the Poisson error
and the uncertainty due to our choice of TRGB threshold and
CMD selection box; however, the true uncertainty
is dominated by systematics, the magnitude of which can be estimated by comparing the
squares and triangles.
\textit{Top-right inset:}
The points indicate the locations of RGB sources in the IMACS field, projected onto
the sky at the distance of NGC~253.
The coordinate system is labeled in kpc from the center of the galaxy,
with north up and east to the left.
The field is divided into radial annuli of width $4$~kpc, and further
subdivided into sectors with azimuthal extent $\Delta\theta=20\degr$, as indicated
by the coloring. The squares indicate the locations of the GHOSTS fields.
\textit{Bottom-left inset:}
Azimuthal profiles of the projected TRGB stellar density.
The $x$ axis gives the azimuthal angle, $\theta$, in degrees.
Each line corresponds to a different radius, starting at the top and going down.
The points are colored the same as the corresponding sectors in the
top-right inset.
The triangles and squares are as in the main panel.
Azimuth increases counter-clockwise in the top-right inset, so the right-most
angular sector in this panel is the one closest to the disk plane.}
\end{figure}

The IMACS field was divided into radial bins of width $4$~kpc, excluding the region
within $5$~kpc of the disk plane in order to avoid thin and thick disk stars. These
annuli were further subdivided into angular sectors of azimuthal extents
$\Delta\theta = 20\degr$, in order to determine
the structure of the halo at a given radius
(different azimuthal bin sizes ranging from $16\degr$ to $40\degr$ were also tested,
with no significant differences in our conclusions).
The top-right inset in Figure~\ref{fig:radsectors} shows the sectors;
azimuth is defined with $\theta=0\degr$ along the major axis to the north-east
(upper-left in the Figure) and increasing counter-clockwise.

The radial projected density profile is shown in
the main panel of Figure~\ref{fig:radsectors}.
The error bars are the sum, in quadrature, of the Poisson error in the
number of detected stars in each bin and the $20\%$ uncertainty due to
our choice of TRGB threshold and CMD selection box (see Section~\ref{sec:densitycalib}).
The profile is well fit
by a broken power-law, of inner slope $n=-1.0 \pm 0.4$ and outer slope $n=-4.0 \pm 0.8$
(where the quoted error is the formal uncertainty in the fit, which is
an underestimate; see below),
with a break at $14.9$~kpc.  As demonstrated later,
the break between these power-laws is not due to an intrinsic change in the distribution
of halo stars, but is rather an artifact of using circular annuli to describe a
distribution that is intrinsically flattened. The IMACS counts within the GHOSTS
fields are shown as the triangles, while the HST/ACS counts are shown as the squares.
The difference between the triangles and squares gives an indication of the systematic
uncertainty, $\sim 0.2$~dex, which dominates over the statistical errors shown
in the profiles.
This is discussed in more detail in Section~\ref{sec:ghostsfieldcomparison}.
Given the size of this systematic uncertainty, a more realistic error for the
slope of the radial power law can be obtained by shifting the inner points up by $0.2$~dex;
doing this changes the slope of the inner power law to $n=-1.6$ and the the slope
of the outer power law to $n=-5.3$.
Although it appears that the IMACS counts within the outer GHOSTS field (Field 10)
are much higher than the counts at adjacent radii, this is because the
radial profile averages over all azimuths, while the GHOSTS field is at one
specific azimuth. Indeed, as shown by the azimuthal profiles in the bottom-left inset of
Figure~\ref{fig:radsectors}
(also shown as the gray lines with error bars on the main plot; these
gray points are offset horizontally with azimuth for clarity),
there is a strong trend in density with
azimuth for each radial bin. The IMACS counts in the GHOSTS field are fully
consistent with the counts in the same radial bin when this trend is
taken into account.
Overall, the almost uniform positive gradient in the azimuthal profiles indicate
that the counts towards the galactic plane (at larger azimuth, which is defined to increase
counter-clockwise) are significantly higher than
along the galactic minor axis at the same radius. This is strong evidence that the
halo profile is not spherically-symmetric, but is rather flattened in the direction
of the disk. Another possibility is that we are not seeing a true stellar halo, but
rather a very extended thick disk, which would necessarily have a very flattened
distribution (see discussion in Section~\ref{sec:vertprofile} below).

\subsubsection{Elliptical Radial Profile}%
\label{sec:ellipprofile}

\begin{figure}
\plotone{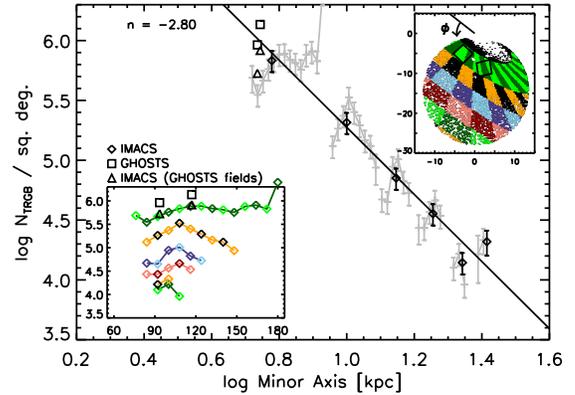}
\caption{\label{fig:ellipsectors}%
As in Figure~\ref{fig:radsectors}, but using elliptical annuli
of axis ratio $b/a=0.35$, with minor
axis widths of $4$~kpc and elliptical sectors with azimuthal extents of $\Delta\phi=8\degr$.
A single power law is fit to the radial profile.
The $x$ axis of the bottom-left inset gives the angular coordinate, $\phi$, in degrees.}
\end{figure}

The azimuthal trend of the halo density in Section~\ref{sec:radprofile} strongly
indicates that the stellar halo of NGC~253 is flattened. Therefore, in order to
determine an accurate radial profile, we must adopt ``annuli'' that mimic the
intrinsic distribution of the stars. In this section, we adopt elliptical annuli
aligned with the stellar disk. For the elliptical axis ratio, we have tested values
ranging from the observed disk axis ratio of $b/a=0.25$
\citep{pence80,koribalski-etal04}, up to $b/a=0.5$.
In the azimuthal direction, we adopt an angular coordinate $\phi$ defined by
\begin{equation}
  \tan \phi = \frac{r_{\mathrm{maj}}}{r_{\mathrm{min}}} \frac{b}{a},
\end{equation}
where $r_{\mathrm{maj}}$ and $r_{\mathrm{min}}$ are the projections of the radial vector
along the major and minor axes respectively. If the halo distribution were perfectly
disk-like, angular sectors spaced uniformly in $\phi$ would correspond to equal volumes. Our
results are qualitatively unchanged if we instead adopt sectors that are spaced uniformly
in the true projected azimuth $\theta$. The elliptical annuli had minor axes spaced
by $4$~kpc, which were further separated into sectors with azimuthal extent
$\Delta\phi = 8\degr$, as shown in Figure~\ref{fig:ellipsectors}
($\Delta\phi$ values from $4\degr$ to $10\degr$ showed similar behaviour).

The radial halo profile in elliptical annuli
with axis ratios $b/a=0.35$ is shown in the main panel of
Figure~\ref{fig:ellipsectors}. A single power-law with a slope of $-2.8 \pm 0.3$
provides an acceptable fit to the data.
This further demonstrates
that the break in the power law seen in Figure~\ref{fig:radsectors} is
entirely an artifact of the assumed geometry: the circular annuli
coupled with the $5$~kpc inner cutoff probe
different ranges of elliptical annuli as a function of radius.
Shifting the inner points up by $0.2$~dex, as in Section~\ref{sec:radprofile},
steepens the slope to $n=-3.4$.

The azimuthal profiles are shown in the bottom-left inset of Figure~\ref{fig:ellipsectors}.
Unlike with
the circular annuli, the densities do not show a systematic gradient,
again demonstrating
the flattening of the halo. However, the azimuthal profiles are far from flat, and
exhibit a strong peak at intermediate azimuths, $\phi \sim 100\degr$. This could
either indicate that the halo isophotes are smooth but ``boxy'', or could be an
indication of discrete structure within the halo. The fact that the peak in azimuth
is broad, rather than being confined to a single bin, indicates that it is not an
artifact of the chip gap (which runs through many of the bins where the density
peaks, particularly
at large radius) or a manifestation of a thin tidal stream; it must be due to an
extended or smooth feature. The shelf seen in the deep optical photometry of
\citet{malin97} extends in this direction, and it is likely that our
azimuthal peak corresponds to this structure.

The choice of axis ratio $b/a=0.35$ was chosen by eye to minimize the variation
of the azimuthal profiles. Values ranging from the disk ellipticity of
$b/a=0.25$ to the value of $b/a=0.4$ that \citet{davidge10-n253} estimates
for the stellar halo all provide reasonable, though somewhat poorer, descriptions.

\subsubsection{Vertical Profile}\label{sec:vertprofile}
\begin{figure}
\plotone{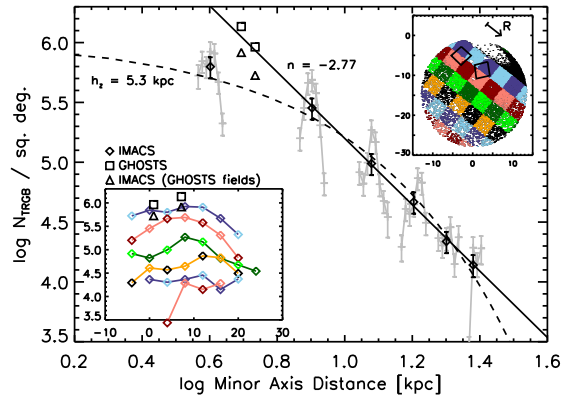}
\caption{\label{fig:rectsectors}%
As in Figures~\ref{fig:radsectors} and \ref{fig:ellipsectors},
for bins of width $4$~kpc perpendicular to the disk and $4$~kpc along the disk
major axis. The $x$-axis in the lower-left inset indicates the distance
along the major axis from the center of the galaxy.
The solid line is a power law fit to all but the innermost point,
which is somewhat compromised by incompleteness within the disk.
The dashed line is an exponential fit to the same data points,
with scale height $h_z = 5.3$~kpc.}
\end{figure}

In the previous section, we found that the geometry of the halo must be quite
flattened, being well fit on ellipses that are almost straight lines parallel
to the disk over much of the field. It is therefore natural to ask whether
a pure disk description, with the density depending only on distance from
the galactic plane, provides a better description. We have constructed a vertical
profile, calculated the density
within vertical slices of $\Delta z=4$~kpc and further subdivided into rectangles
of length $4$~kpc in the direction parallel to the major axis,
as shown in the top-right inset of Figure~\ref{fig:rectsectors}
(major axis subdivisions ranging from $2$ to $8$~kpc gave similar results).
The innermost vertical sheet is somewhat compromised by incompleteness within
the disk near the minor axis, and so the density of this bin was not used
when fitting the density profile.
The shapes of the ellipses and of the vertically-stratified sheets
are not too dissimilar over much of the field,
and so it is not surprising that both the quality and slope ($-2.8 \pm 0.2$,
steepening to $-3.3$ if the inner points are systematically low)
of the best fit power law are similar to the elliptical case.
The density variation parallel to the major axis
within each layer is also similar to the angular variation within the elliptical
annuli.

Given that a description where the density varies primarily in the direction
perpendicular to the disk is acceptable, it is natural to wonder whether the
extended stellar structure we have detected is truly the stellar halo, or is
rather a thick disk.
This is pertinent given that M31, the galaxy whose outer regions are the
best studied, shows evidence for both an extended rotating disk-like
component \citep{ibata-etal05} and a thick disk
of estimated scale height $2.8$~kpc \citep{collins-etal11}.
In order to test this hypothesis, we have fitted a vertical exponential to the
density structure we have detected (again omitting the innermost point),
shown as the dashed line in
Figure~\ref{fig:rectsectors}. An exponential is found to fit the data as
well as the power law, but with an extremely large scale height of $h_z = 5.3 \pm 0.5$~kpc
($4.5$~kpc if the inner points are moved systematically higher).
While we cannot rule out this possibility with the present data, we believe
that a power law halo is a more likely explanation for the following reasons:
\begin{enumerate}
 \item Such a large scale height is significantly larger than that of
other stellar disks of comparable galaxies. While it is only $1.9$ times
larger than the $2.8$~kpc thick disk scale height estimated by \citet{collins-etal11}
for M31, these authors did not directly measure the scale height but rather
assumed that the scale height varies with scale length in the same manner as
the sample of low surface brightness galaxies studied by \citet{yd06}.
The large thick disk scale height inferred for M31 is thus a direct
consequence of its long disk scale length. NGC~253, on the other hand, has
a much smaller scale length; estimates based on surface photometry at
wavelengths from the optical through far infrared place 
the thin disk scale length between
$1.5$ and $2.6$~kpc, once scaled to our adopted distance of $D=3.48$~Mpc
\citep{pence80,fd92,radovich-etal01}.
Adopting a thin-to-thick disk scale length ratio of $1.3$ and the thick disk
scale height-to-scale length ratio of $0.35$ (\citealp{collins-etal11},
based on \citealp{yd06}), the expected thick disk scale height is between
$0.7$ and $1.2$~kpc, many times smaller than what we derive.
The derived scale height would require a very large vertical
velocity dispersion to maintain; for a stellar mass surface density in
the disk of $75~M_{\sun}~\mathrm{pc^{-2}}$
(as for the Milky Way at the solar circle, \citealp{BT08}; this is
a reasonable assumption given the similar total luminosity and disk
scale length of the two galaxies),
$\sigma_z = 104~\mathrm{km~s^{-1}}$, making it essentially a pressure-supported
structure.
 \item Both IMACS (GHOSTS fields) data points lie above the extrapolation of the
exponential fit.
 \item The apparent quality of the exponential fit comes mainly from the innermost
point, but this point is likely an underestimate due to incompleteness.
\end{enumerate}

This issue can be resolved with more data that are deeper and cover a larger
area; new HST GHOSTS data on NGC~253 and other galaxies are being analyzed
to address the outer shape and profile of extended halo components.

\subsection{Density Map}\label{sec:densitymap}

\begin{figure*}
\plotone{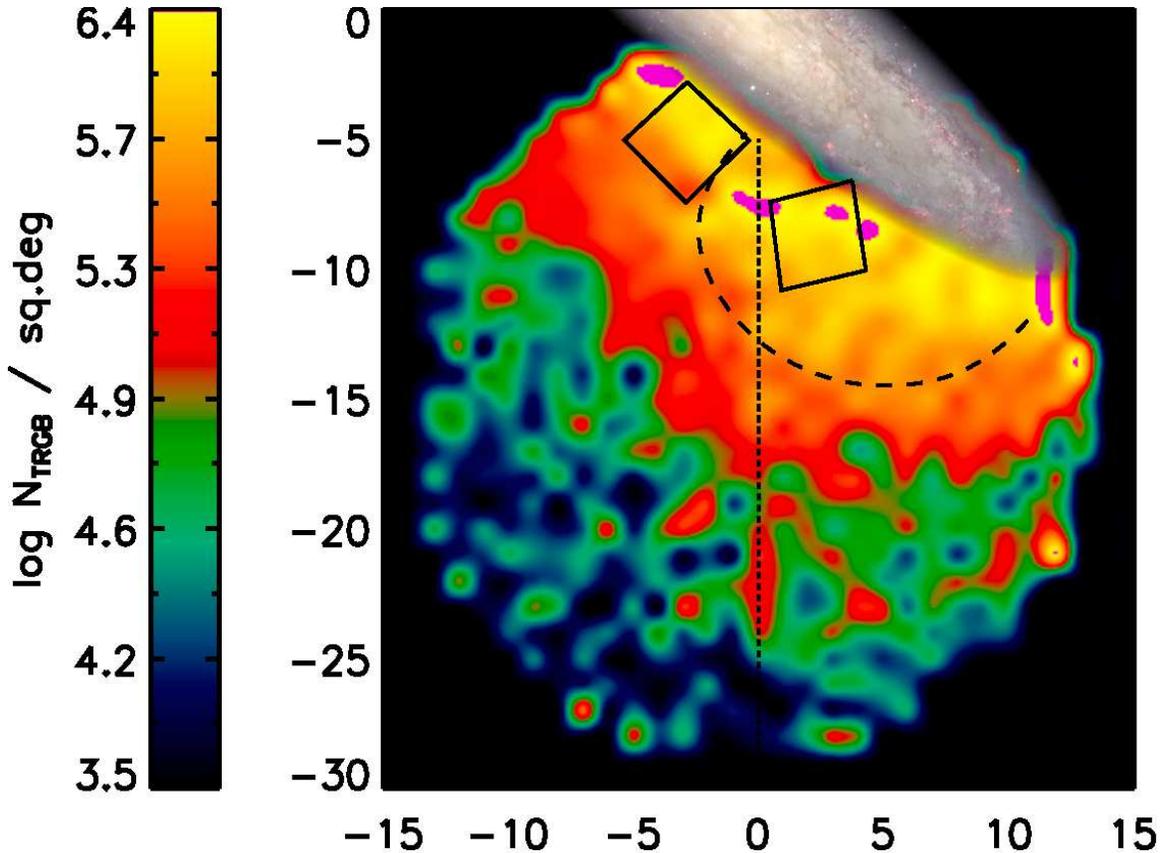}
\caption{\label{fig:densitymap}%
Map of the density of TRGB stars in the halo of NGC~253. The map is oriented with
north up and east to the left. The $x$ and $y$ axes are labeled in kpc from the
center of the galaxy. Magenta indicates where the derived density is a lower limit
on the true density. The GHOSTS pointings are denoted with the black squares,
the location of the shelf is shown by dashed lines, and the vertical dotted line
denotes the horizontal chip gap, where the counts are artificially inflated.
The dark ellipse at the top is the main body of the galaxy, where
individual sources can no longer be distinguished; an optical image of the
galaxy, to scale, has been overlayed for reference (image courtesy of
T.A. Rector/University of Alaska Anchorage, T. Abbott and NOAO/AURA/NSF).
The map was generated using $1~\mathrm{kpc}~\times~1~\mathrm{kpc}$ pixels and
then smoothed using the minimum curvature surface.
Note that the mapping between the color scale and density value has been chosen to
enhance visibility, so attention should be paid to the labels on the color bar,
which give the logarithm of the TRGB number density per square degree.}
\end{figure*}

A map of the derived density of TRGB stars in the halo of NGC~253
is presented in Figure~\ref{fig:densitymap}.
This map was produced by binning the stars into pixels $1$~kpc on a side,
and performing the background subtraction and artificial star correction.
The image was then stretched using histogram equalization to enhance visibility,
smoothed using the minimum curvature surface algorithm \citep{franke82},
and resampled to a resolution of $0.1$~kpc.
The main body of the galaxy,
where we cannot reliably detect stars, is apparent as the dark ellipse at the
top (an optical image of the galaxy is overlayed, for reference),
while the GHOSTS pointings are denoted by the black squares. Magenta
indicates regions where the derived density reaches the saturation level
in the IMACS data and we are therefore only able to place a lower limit on
the true stellar density at these points.

A few features are notable in this map. First and foremost, the stellar density
falls off strongly with radius from the galaxy, demonstrating that the stars belong
to NGC~253. Secondly, the isopleths of constant stellar density are ellipses
whose axis ratio is similar to or somewhat larger than
that of the observed galaxy, as anticipated from
Section~\ref{sec:ellipprofile}. Thirdly, there are are some large-scale
substructures, chiefly a shelf-like feature near the center of the field
(shown schematically by the dashed curve; see also \citealt{malin97} figure 4).
Finally, there is a small scale pixel-to-pixel variation in the density.
None of these features appears to be associated with any overdensity of
resolved background galaxies.
Note that the vertical feature at
$x=0$ extending from $y=-20$ to $-25$~kpc lies right along the
chip gap (shown with the dotted line), and is therefore most likely an artifact.

\begin{figure*}
\center{
\includegraphics[height=1.7in]{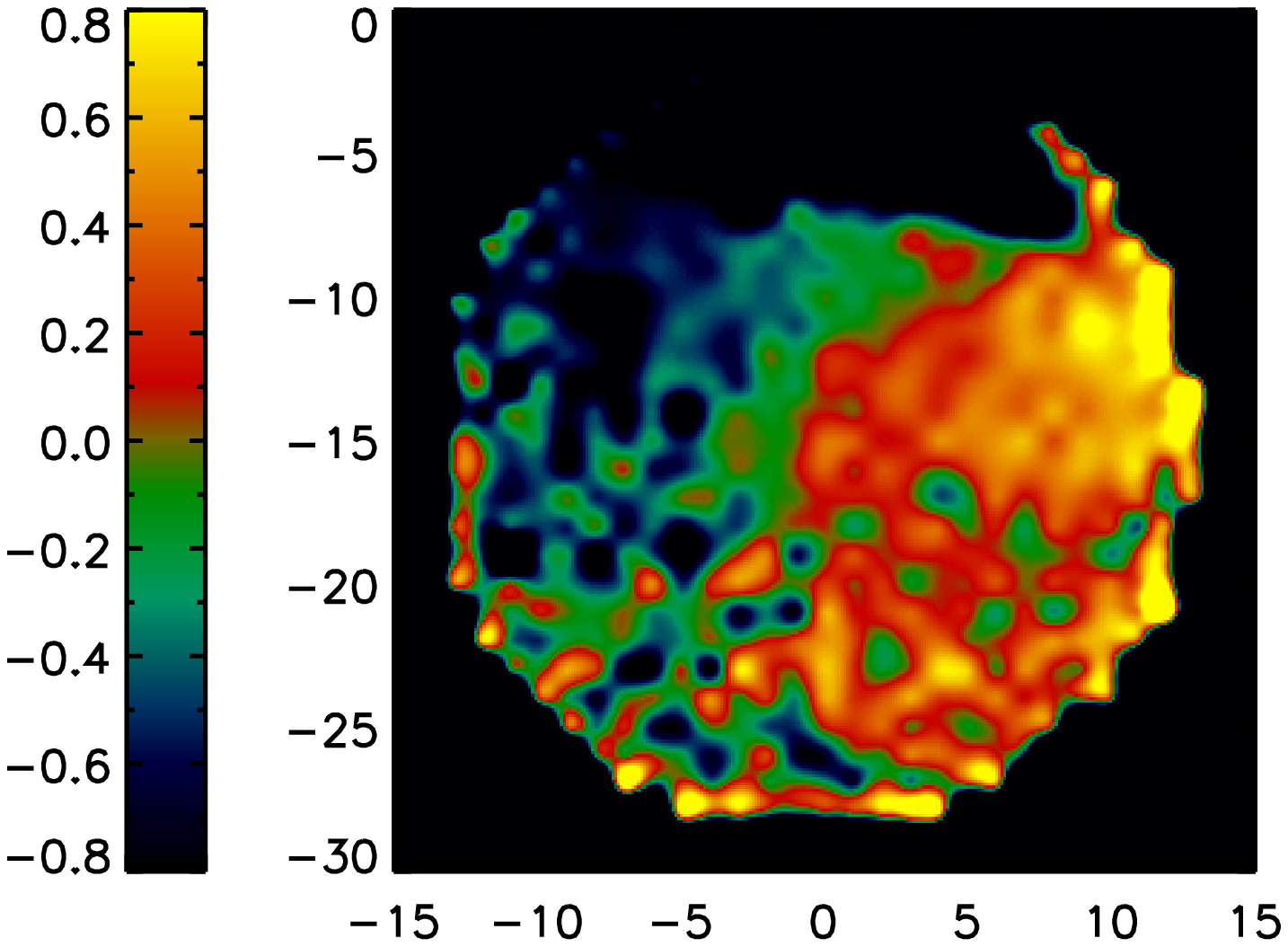}
\includegraphics[height=1.7in]{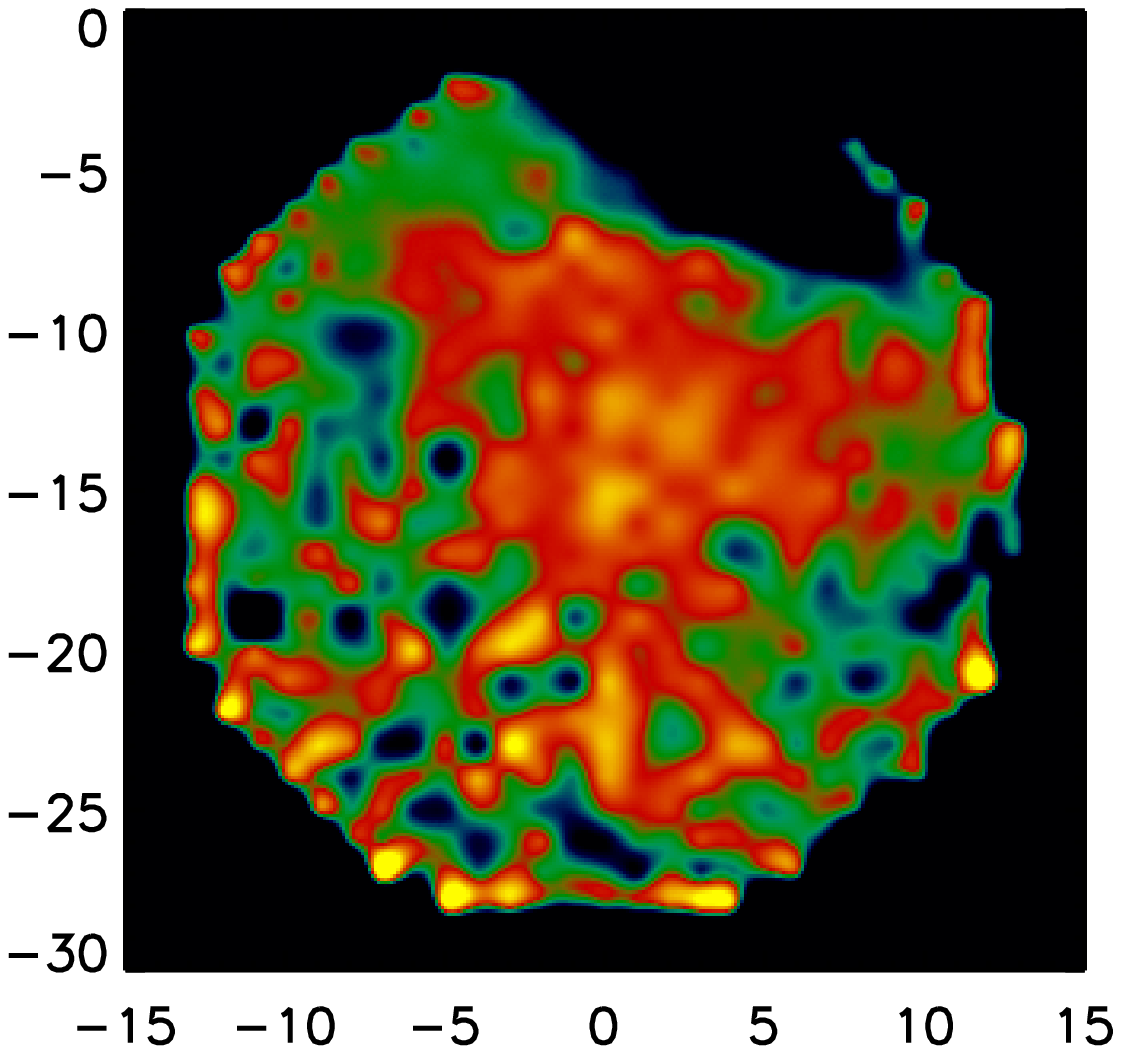}
\includegraphics[height=1.7in]{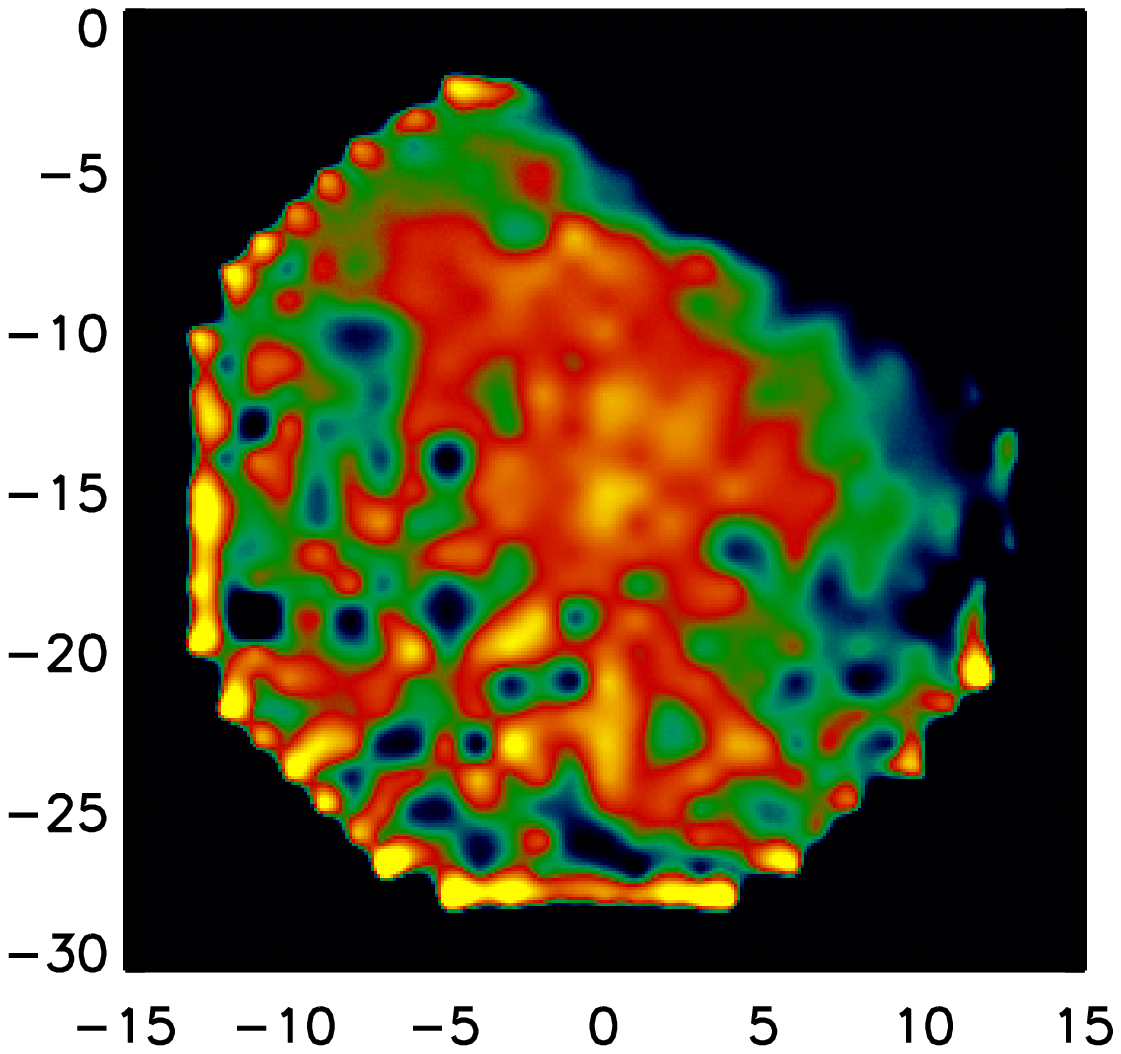}
}
\caption{\label{fig:densitymapoverpow}%
Ratio between the density map shown in Figure~\ref{fig:densitymap} and the
best fit radial \textit{(left)}, elliptical \textit{(middle)} and vertical
\textit{(right)} power law from Figures~\ref{fig:radsectors}, \ref{fig:ellipsectors},
and \ref{fig:rectsectors} respectively. The color scale label is the logarithm of
the ratio. Axes are labeled in kpc from the center of the galaxy.}
\end{figure*}

In Figure~\ref{fig:densitymapoverpow}, we show the residuals between the
density map and the best fit power laws, expressed as the logarithm of the
ratio. The left-hand panel shows the result using the radial power law.
The systematic trend for the residuals to be negative on one side and
positive on the other side demonstrates again that a simple radial power law
is a poor description of the halo, i.e.~it is not spherically symmetric.
In the middle panel, the elliptical power law fit is used. In this case, 
the residuals show no systematic trend with azimuthal angle, demonstrating that
the halo is consistent with being flattened in the same sense as the disk.
The vertical feature at $x=0$, which is an artifact of the chip gap, is
very apparent, along with smaller-scale variation.
In the right-hand panel, the vertical power law fit is used; given the
similarity between the elliptical and vertical geometry, it is not surprising
that this panel shows similar features to the middle panel.
However, there is a small systematic negative trend at the on the right side of the image,
which again suggests that a power law halo is a better description than an
exponential thick disk.

\subsection{GHOSTS Fields}%
\label{sec:ghostsfieldcomparison}

The overlap between the GHOSTS fields and the IMACS data provides a check
on our methodology. The density of TRGB stars
brighter than $I<24.25$ in the HST/ACS GHOSTS data is
$\log N_{\mathrm{TRGB}} = 5.96$ and $\log N_{\mathrm{TRGB}} = 6.13$
for Fields 9 and 10 respectively,
while the derived densities from the IMACS data over the identical areas of sky
are $\log N_{\mathrm{TRGB}} = 5.73$ and $\log N_{\mathrm{TRGB}} = 5.92$,
factors of $1.7$ and $1.6$ lower, respectively.
While these factors are not unity, they are reassuringly close given the
dramatic difference in image quality and depth between the HST/ACS and IMACS
data, and validates the
use of moderate ground-based data to study stellar halos of galaxies
beyond the Local Group.

The factor of $\sim 1.7$ ($\sim 0.2$~dex) provides an indication of the magnitude of
uncertainty that is introduced by using ground-based data. Because the factor is
similar in both fields, it is not a random error but must reflect a systematic
uncertainty, and dominates over the random Poisson error (which is denoted
by the error bars of Figures~\ref{fig:radsectors} -- \ref{fig:rectsectors}).
With only two overlapping HST fields, which are both located at similar densities,
we cannot determine whether this systematic reflects an overall normalization
error, or is specific to the highest density regions. Overlapping HST data at
lower stellar densities will be required to disentangle these possible effects.
Although there is a small magenta region in Figure~\ref{fig:densitymap} within
the GHOSTS Field 10 pointing, indicating that the IMACS data give only a lower limit at
that location, the lack of magenta within the Field 9 pointing indicates that this
is not the cause of the offset.

Note that although the photometric accuracy of the IMACS data is insufficient
to determine metallicity via the color of the RGB, 
\citet{ghosts11} report an
[Fe/H] estimate based on the TRGB colors of $-1.1 \pm 0.3$ for the HST/ACS
GHOSTS data in both of these fields.

\subsection{Shelf}\label{sec:shelf}
As noted in Section~\ref{sec:densitymap}, an overdensity of
stars is seen on the south side of the galaxy. A rough outline of the
region is shown with the dashed curve in Figure~\ref{fig:densitymap}. This
overdensity coincides with the optical shelf seen in deep images of
\citet{malin97}. Our detection of this feature demonstrates that it
is due to stars, rather than synchrotron emission \citep[as suggested by][]{beck-etal82}.

Knowledge of the stellar content in the shelf would greatly aid in understanding
its origin. Unfortunately, the high density in this region results in significant
blending in the IMACS data, and therefore errors in the stellar colors that are
too large to allow the detection of stellar population differences. However, the locations of the
GHOSTS pointings are fortuitous: Field 10 lies on the shelf, while Field 9 lies just
off the shelf. These fields are at almost identical elliptical radii, so we would
expect the stellar populations in these fields to be very similar if the halo
were smooth. We can therefore compare the colors of stars detected in the
HST/ACS data in these fields to constrain the stellar population of the shelf.

\begin{figure}
\plotone{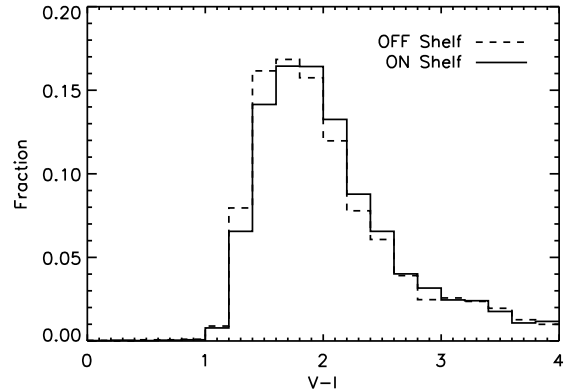}
\caption{\label{fig:shelfcolourhist}%
Histograms of the $V-I$ color distribution of stars at the TRGB
in GHOSTS Fields 9 (``OFF Shelf'') and 10 (``ON Shelf'').}
\end{figure}

We have plotted the fraction of TRGB stars ($I < 24.25$) as a function of
their $V-I$ color for the two GHOSTS fields in Figure~\ref{fig:shelfcolourhist}.
The distribution is very similar, indicating that the stellar populations
in the shelf are not dramatically different from the rest of the halo at
this distance.
The differences that are seen are in the sense that the shelf contains more red
($1.8 < V-I < 2.6$) TRGB stars, and fewer
blue ($1.2 < V-I < 1.8$) TRGB stars. Using the transformation
between metallicity and TRGB color
in \citet{bellazzini-etal01}, this corresponds to more stars at
$-1.3 < \mathrm{[Fe/H]} < -0.7$ and fewer at $\mathrm{[Fe/H]} < -1.3$.
Indeed, using the same data, \citet{ghosts11} find that the Field 10 TRGB
is $0.06$~mag redder and $0.05$~dex more metal-rich than Field 9.

A higher metallicity in the shelf is consistent with
hierarchical models of stellar halo formation, which predict that the highest surface
brightness substructures, having been formed from the largest satellites,
are more metal-rich than the well-mixed smooth component \citep{font-etal08}.

\section{Halo Luminosity and Mass}%
\label{sec:haloluminosity}

\subsection{Synthetic CMD Calibration}
\label{sec:iacstar}
In order to translate the number of TRGB stars to a total luminosity
of the underlying stellar population, we used the online IAC-STAR%
\footnote{http://iac-star.iac.es/iac-star/} synthetic
CMD code \citep{iacstar} to generate an old metal-poor stellar population,
and then compared the number of TRGB stars to the total luminosity.
We tested several sets of parameters for the stellar population: a single
burst lasting for $1$~Gyr that ended $12$~Gyr or $9$~Gyr ago, and a
metallicity of $Z=0.0007 \approx 1/30~Z_{\sun}$, $Z=0.002 \approx 1/10~Z_{\sun}$,
or $Z=0.007 \approx 1/3~Z_{\sun}$,
in all cases using a \citet{Kroupa93} initial mass function,
the \citet{girardi-etal00} stellar evolution library,
and bolometric corrections based on the \citet{ck04} models.
The synthetic CMDs contained between $10000$ and $20000$
stars above $M_I < -3.46$ (equivalent to our TRGB cut at an observed magnitude
of $I < 24.25$).
The resulting stellar populations had a total integrated luminosity
per TRGB star of between
$1.2 \times 10^4$ and $1.5 \times 10^4~L_{\sun}/N_{\mathrm{TRGB}}$,
increasingly monotonically with metallicity but having very little dependence
on the age of the population ($L/N_{\mathrm{TRGB}}$ is $\approx3\%$ higher for
the older population).
We adopt a scaling of $1.3 \times 10^4~L_{\sun}/N_{\mathrm{TRGB}}$,
appropriate for an old population with metallicity $Z = 0.1~Z_{\sun}$,
close to the inferred metallicity in the overlapping GHOSTS fields \citep{ghosts11},
and caution that large differences between the assumed and intrinsic
metallicities could introduce $\sim 20\%$ systematic errors in the
derived luminosities.

Translation to total stellar mass is more problematic due to a numerical
issue in the IAC-STAR code that, for some sets of input parameters,
gives a NaN result for some integrated quantities of the stellar population,
such as extant stellar mass (A.~Aparicio 2010, private communication).
Using the stellar mass when available,
and extrapolating those mass-to-light ratio trends with metallicity and age
to those sets of parameters for which the stellar mass
was unavailable, we find that the total stellar mass per TRGB star is between
$1.2 \times 10^4$ and $2.1 \times 10^4~M_{\sun}/N_{\mathrm{TRGB}}$, increasing
as a function of both metallicity and age.
We adopt a scaling of $1.6 \times 10^4~M_{\sun}/N_{\mathrm{TRGB}}$,
and again caution that different stellar populations could introduce
systematic errors in the derived masses of order $\sim 50\%$.

\subsection{Halo luminosity and mass}

The total number of TRGB stars we derive in our field is
$N_{\mathrm{TRGB}} = 3.5 \times 10^4$, resulting
in a minimum halo luminosity of $L_{\mathrm{halo}} > 4.6 \times 10^8~L_{\sun}$ and minimum
halo mass of $M_{\mathrm{halo}} > 5.6 \times 10^8~M_{\sun}$. We can estimate
the total halo luminosity several ways. First, we note that we have covered
approximately $30\%$ of the area of the galaxy to a distance of $30$~kpc,
and that the halo density drops sufficiently steeply that the majority of
the stars should be contained within $30$~kpc, resulting in a total
of $L_{\mathrm{halo}} = 1.5 \times 10^9~L_{\sun}$ or
$M_{\mathrm{halo}} = 1.9 \times 10^9~M_{\sun}$.
Another estimate of the total luminosity
can be determined from the elliptical power law fit to the TRGB radial number density;
however, as the slope of the projected density is significantly
more negative than $-1$, the integral
diverges at small radius. We have therefore imposed an inner cutoff at a
radius along the minor axis of $5$~kpc,
and caution that this estimate does not account for most of the halo stars that
overlap with the galactic disk.
The total halo luminosity using these assumptions
is $L_{\mathrm{halo}} = 2.6\pm0.3 \times 10^9~L_{\sun}$,
while the halo mass is $M_{\mathrm{halo}} = 3.2\pm0.4 \times 10^9~M_{\sun}$.
The main difference between these two estimates is due to small-scale structure,
rather than the counts beyond $30$~kpc which only account for $15\%$ of the luminosity
using the power law fit.
The quoted random errors, which are due to the uncertainty in the power law fit,
certainly underestimate the true uncertainty; for example,
systematically shifting the inner points by the estimated systematic uncertainty
of $0.2$~dex, as in Section~\ref{sec:profiles},
changes the luminosity and mass by $50\%$.
The confluence of these various estimates suggests that the total halo
luminosity is $L_{\mathrm{halo}} \approx 2 \pm 1 \times 10^9~L_{\sun}$ and the stellar
mass is $M_{\mathrm{halo}} \approx 2.5 \pm 1.5 \times 10^9~M_{\sun}$.

We estimate the fraction of the total light and stellar mass of the galaxy
contained in the halo from the total RC3 \citep{RC3} $B$ and $V$ magnitudes
of $B=8.04$, and $V=7.19$, which gives an absolute magnitude of $M_V=-20.6$
and a $V$-band luminosity of $\approx 1.5 \times 10^{10}~L_{\sun}$. This corresponds
to a total stellar mass of $\approx 4.4 \times 10^{10}~M_{\sun}$, assuming
$M/L_V=2.9$, as given by the $B-V$ vs. $M/L_V$ relation in \citet{bdj01}.
Therefore, a fraction
$\sim 0.13$ of the luminosity and $\sim 0.06$ of the stellar mass of NGC~253 is in the
form of its halo. This is consistent with theoretical models that predict that the median
galaxy at this luminosity contains a stellar halo with $1$ -- $5\%$ of
the galaxy stellar mass \citep{purcell-etal07}.

The only other galaxies for which the total luminosity and mass of the stellar
halo can be obtained are the Milky Way and M31, both of which are of similar luminosity to
NGC~253. Measurements place the total luminosity and mass of the Milky Way's stellar
halo at $L_V \sim 10^9~L_{\sun}$ and $M \sim 2 \times 10^9~M_{\sun}$ respectively
\citep[see the summary in][]{bj05},
and the luminosity of the M31 stellar halo at $L \sim 10^9~L_{\sun}$ \citep{ibata-etal07},
similar in magnitude to but somewhat smaller than that of NGC~253.

\section{Conclusions}
We have detected resolved RGB stars in the stellar halo of NGC~253 in both
HST/ACS and ground-based Magellan/IMACS imaging data. The clean HST data
are only able to probe a small area of the halo, but provide a benchmark
by which we can evaluate the ground-based data. The large area covered by
a single IMACS pointing provides a global picture of the halo, reaching
out to distances of $30$~kpc from the center of the galaxy, which is
required given the substructure and large-scale features predicted in models
of stellar halo formation and seen in the observations.

While previous studies of resolved stellar halos have successfully detected stars
in the halo
and discovered spectacular tidal features
\citep{mouhcine-etal10}, only for the
Milky Way and M31 has there been significant quantitative analysis of the halo profile
and structure. In this work, we have performed detailed artificial star
tests in order to make quantitative
measurements that can be compared to theoretical predictions and compared
to observations of other galaxies.

The projected stellar density declines with
elliptical radius as a power law of slope $-2.8 \pm 0.6$. This is consistent with,
but at the steep end, of the density profiles found for the Milky Way and M31 halos;
for example, \citet{bell-etal08} find that the three-dimensional density
falls off as a power of index of between $-2$ and $-4$, corresponding to projected radial
power law indices of $-1$ to $-3$,
and the projected radial profile of the halo of M31 
has a power law index of $\approx -2$ \citep{ibata-etal07}.
An exponential vertical distribution is
also found to provide an acceptable fit, with an
extraordinarily large scale height of $h_z = 5.3 \pm 0.8$~kpc.
However, interpretation of the observed structure as a thick disk rather
than a power law halo would require an unrealistically large vertical velocity
dispersion, and is inconsistent with the observed halo density
in the HST/GHOSTS fields.

The halo is found to be flattened in the same direction as the stellar disk,
with a projected axis ratio of $\approx 0.35 \pm 0.1$. This is consistent with what
has been found for other galaxies:
the Milky Way halo is also flattened in the same sense as its disk at small radii
\citep[e.g.][]{carollo-etal07},
\citet{dejong-etal09} find a typical halo axis ratio of $c/a \approx 0.4$ based
on HST/GHOSTS data of nearly edge-on galaxies,
and although not quantitative, the star counts in the halo of
NGC~891 also appear significantly flattened like the disk \citep{mouhcine-etal10}.
The dark matter halo that dominates
the potential is also predicted in galaxy formation simulations
to be flattened in the same sense as the galactic disk within $10\%$ of the
virial radius \citep{bailin-etal05-diskhalo}.
Observations of more galaxies are required to see if this is
a universal feature of stellar halos.

\citet{davidge10-n253}, who detected AGB stars in the halo of NGC~253,
found their distribution to be flattened similarly to what is found in this study
for the older RGB stars. Unfortunately, our observations are not sensitive
to these younger stars, and so further observations will be
required in order to fully understand the relationship between these populations.

Density maps of the halo structure show substructure on a variety of scales.
The shelf-like feature seen in earlier deep photographic studies
\citep{malin97} is evident.
In addition, there is a significant amount of smaller kpc-scale structure.
These features are not apparent in the small HST fields but require the full
IMACS field of view to appreciate.

We estimate the total luminosity of the stellar halo to be $\sim 2\pm1 \times 10^9~L_{\sun}$,
and the stellar mass to be $\sim 2.5\pm1.5 \times 10^9~M_{\sun}$. This corresponds to
a fraction $0.13$ of the total luminosity and $0.06$ of the total stellar mass
of the galaxy. These values are consistent with the theoretical models of
\citet{purcell-etal07}.
The NGC~253 halo luminosity is a factor of $\approx 2$ larger than the halos of
the Milky Way and M31, despite the similar total luminosity of these galaxies,
indicating significant galaxy-to-galaxy scatter in the relative halo fraction,
as predicted by models.

This paper provides a roadmap for analyzing ground-based imaging data of the
stellar halos of galaxies beyond the Local Group in a quantitative way.
These techniques will be applied in future work to a series of other galaxies
for which we have obtained similar data. The resulting collection of
quantitative global measurements of the scale and structure of stellar
halos of a broad sample of galaxies will allow for statistical comparison
to models of stellar halo formation, and will provide vital constraints
on the galaxy assembly processes that are traced by the halos.

A catalog of the detected sources is being made available as
a table on the Centre de Donn\'ees astronomiques de Strasbourg (CDS).

\acknowledgments
We thank Anna Gallazzi for helping obtain the Magellan observations and
Chien Peng for useful data reduction advice.
This work has made use of the IAC-STAR Synthetic CMD computation code. 
IAC-STAR is supported and maintained by the computer division of the IAC.
This research has made use of the NASA/IPAC Extragalactic Database (NED) which is operated by the Jet Propulsion Laboratory, California Institute of Technology, under contract with the National Aeronautics and Space Administration.
This work was supported by the National Science Foundation through grant AST-1008342,
and by HST grant GO-11613. Support for Program number GO-11613 was provided by NASA
through a grant from the Space Telescope Science Institute, which is operated by the
Association of Universities for Research in Astronomy, Incorporated, under
NASA contract NAS5-26555.

{\it Facilities:} \facility{Magellan:Baade (IMACS)}, \facility{HST (ACS)}

\bibliography{../../masterref.bib}

\end{document}